# Weak antilocalization and ferromagnetism in magnetic Weyl semimetal $Co_3Sn_2S_2$

Kapil Kumar[1,2], M. M. Sharma[1,2] and V.P.S. Awana[1,2*]

[1]*CSIR-National Physical Laboratory, Dr. K. S. Krishnan Marg, New Delhi-110012, India.*

[2]*Academy of Scientific and Innovative Research (AcSIR), Ghaziabad 201002, India.*

**Abstract:**

Here we report synthesis of single crystalline magnetic Weyl semimetal $Co_3Sn_2S_2$. The synthesized crystal is characterized through various tools viz. X-ray diffraction, field emission electron microscopy and X-ray photoelectron spectroscopy. A clear ferromagnetic transition is observed in magnetization and heat capacity at around 175K, which is further verified through electrical transport measurements. Hysteresis is observed in ρ-T measurements in cooling and warming cycle, showing the presence of first order phase transition and charge ordering in the synthesized sample. The synthesized $Co_3Sn_2S_2$ exhibits high magnetoresistance of around 230% at 2K. The transport phenomenon in synthesized $Co_3Sn_2S_2$ appears to have contributions from topological surface states at low temperature below say 70K, and above that the same is found to be strongly dependent on its bulk magnetic state. Magnetoconductivity data at low fields of upto ±1T (Tesla) is fitted with Hikami Larkin Nagaoka model, which shows the presence of weak antilocalization effect in synthesized $Co_3Sn_2S_2$ crystal at low temperatures below 30K. Angle dependent magneto transport measurements confirms that the observed WAL is the topological surface states dominated phenomenon.

[*]**Corresponding Author**

Dr. V. P. S. Awana:  E-mail: awana@nplindia.org

Ph. +91-11-45609357, Fax-+91-11-45609310

Homepage: awanavps.webs.com

**Introduction:**

Magnetic topological materials are recently emerged as a hot topic of research in condensed matter physics. The presence of magnetic element in topological materials enables one to play with both the topological surface and beneath bulk electronic states, and thus creating the possibility to realize quantum anomalous Hall effect (QAHE). The topological materials with intrinsic magnetic ordering are considered to be the best suited materials for QAHE [1-3]. Some of the materials viz. $MnBi_2Te_4$ [2,4-6], $Co_3Sn_2S_2$ [7-13] are found to show intrinsic magnetic ordering with topological non trivial band structure. $Co_3Sn_2S_2$ comes under the class of ternary Shandite compounds with general formula $T_3M_2X_2$ (T =Ni, Co, Rh or Pd; M = Sn, In, or Pb; and X = S or Se) [14]. The bulk electronic band structure of $Co_3Sn_2S_2$ resembles with Weyl semimetals (WSMs) [15,16]. WSMs are realized by breaking of time reversal symmetry (TRS) [17], inversion symmetry [18] or both [19]. The bulk and surface fermi arcs of WSMs connects the Weyl points of opposite chirality and hosts Weyl fermions [17]. The WSMs which emerges due to breaking of TRS contains two Weyl points on the other hand the ones having broken inversion symmetry have at least four Weyl points [17].

$Co_3Sn_2S_2$ is considered as TRS breaking WSM and extensively studied in recent years due to its remarkable properties such as chiral anomaly [7], giant anomalous Hall conductivity [9], non-saturating high MR [7], 1-D chiral edge states [20], and planar Hall effect [21]. $Co_3Sn_2S_2$ show ferromagnetic ordering at around 175K [7-13]. $Co_3Sn_2S_2$ crystallizes with a 3-2-2 like kagome structure with kagome layers of Co-Sn being sandwiched between S atoms [22]. The materials with kagome structure are found to show a variety of interesting magnetic properties along with topological non trivial band structure in some cases [23-26]. $Fe_3Sn_2$ [23], $Mn_3Sn$ [24], CoSn [25], $TbMn_6Sn_6$ [26] etc. are the examples of such materials having intrinsic magnetic moment with non-trivial band topology. The simultaneous existence of magnetic ordering and non-trivial topological structure makes the topological kagome materials a suitable choice for realization of AQHE [27,28]. In this regard, $Co_3Sn_2S_2$ is reported to show huge anomalous Hall effect (AHE) due to its large Berry curvature [9]. The observed large AHE is found to be robust against magnetic doping [29,30]. $Co_3Sn_2S_2$ also shows non saturating high magnetoresistance (MR) similar to as many other topological semimetals do [7,9,21,30]. Interestingly, till now, $Co_3Sn_2S_2$ is not studied in the context of weak antilocalization (WAL) effect, which is considered as the vital property of the topological materials. In this article we touch upon this point by performing the isothermal MR measurements at various temperatures below the magnetic ordering temperature. The MR in different temperatures follows varying trends and more interestingly the same at low temperatures i.e., deep below the magnetic ordering is clearly dominated by the surface states as envisaged by WAL effect. For this study, a single crystal of $Co_3Sn_2S_2$ is synthesized by following a multi-step solid state reaction route. Intrinsic ferromagnetic ordering at around 175K is verified through electrical transport, magnetization and heat capacity measurements. The low field magneto conductivity (MC) of $Co_3Sn_2S_2$ is modelled with HLN formalism, and it shows the presence of WAL effect in the synthesized sample. Though, the magneto transport of $Co_3Sn_2S_2$ is extensively studied in recent years, yet the presence of WAL effect in the same is remained unexplored till date, which is studied in the present article.

**Experimental:**

Single crystals of $Co_3Sn_2S_2$ were grown using a 3-step solid state reaction route as reported in ref. [7]. High purity powders of Co, Sn, and S were weighed in stoichiometric ratio (3:2:2) and then mixed and grounded thoroughly in Argon atmosphere inside MBRAUN glove box. Thus, obtained homogenous mixture was palletized by hydraulic press and vacuum encapsulated in quartz ampoule. The ampoule was first preheated to 500°C for 24 hours and cooled down to room temperature. This pre heated sample is further heated to 700°C for 24 hours for phase formation. In the last step, the obtained powder of $Co_3Sn_2S_2$ is melted at high temperature, 950ºC for 24 hours and then slowly cooled to 800ºC at a rate of 1ºC/h. The sample is dwelled at 800ºC for 24 hours to stabilize the phase and then normally cooled to room temperature. The schematic of complete heat treatment along with the image of synthesized crystal is shown in Fig. 1. Thus, obtained silvery shiny crystals have the dimension 7 mm×3 mm×2 mm and are easily cleavable along the growth axis.

Rigaku made table-top Miniflex-II X-ray diffractometer was used to record XRD pattern on gently crushed powder and mechanically cleaved crystal flake. Rietveld refinement of powder XRD pattern (PXRD) was obtained using FullProf software. Crystallographic information file (CIF) obtained from Rietveld refinement is processed in VESTA software to draw the unit cell. Zeiss EVO-50 made Field emission Scanning electron microscopy (FESEM) equipped with energy dispersive X-ray analysis (EDAX) was used to visualize surface morphology and elemental composition, respectively. Thermo Fisher made XPS spectrometer is used to record XPS spectra. A vacuum of $7\times10^{-9}$ torr was maintained during the measurement and Al K-α radiation was used as source. Electrical transport measurements and thermal measurements were performed using Quantum Design Physical Property Measurement System (QD-PPMS). Magnetization measurements were performed using Quantum design Magnetic Properties Measurement System (QD-MPMS).

**Results & Discussion:**

Fig. 2(a) exhibits the XRD pattern, being taken from the surface of mechanically cleaved crystal flake of $Co_3Sn_2S_2$ single crystal. High intensity peaks are observed for some specific $2\theta$ values, which corresponds to (003), (006), (009) and (0012) planes. This signifies that the synthesized crystal is grown unidirectionally along c axis and the planes are stacked in (003n) manner. Full width at half maxima (FWHM) of the peak along (006) planes (peak with highest intensity) is found to be 0.20°, which shows highly crystalline nature of the synthesized $Co_3Sn_2S_2$ crystal. Phase purity of synthesized $Co_3Sn_2S_2$ crystal is verified through Rietveld refined PXRD pattern, which is shown in fig. 2(b). PXRD pattern is well fitted with applied parameters of a rhombohedral crystal structure with R -3 m space group. All peaks are indexed and fitted within their permitted respective planes and unidentified species are not seen within the XRD limit. The parameter of quality of fit i.e., $\chi^2$ is found to be 1.79, which is reasonably good. The unit cell parameters obtained from Rietveld refinement along with the fitting parameters are listed in table 1. The CIF generated from Rietveld refinement is utilized to draw the unit cell structure in VESTA software, and the same is shown in fig. 2(c). It is clearly shown in unit cell structure that Co-Sn layers forms the kagome structure, which are sandwiched between the S atoms. Also, there are total three such arrangements in a single unit cell. This is in agreement with the observed (003n) peaks seen in XRD pattern of mechanically cleaved crystal flake surface.

Fig. 3(a) shows FESEM image taken on flake of synthesized $Co_3Sn_2S_2$ single crystal. FESEM image is showing a layered terrace type morphology, which shows the crystalline nature of the synthesized sample. Also, no colour contrast has been observed in FESEM image, which shows that the crystal is grown in single phase. EDS spectra is recorded for elemental analysis and the same is shown in fig. 3(b). EDS spectra shows the presence of peaks of all constituent elements viz. Co, Sn and S. No extra peak has been observed for any foreign impurity element, which signifies the purity of the sample. The elemental composition is also determined, which is given in inset of fig. 3(b). All constituent elements are found to be near to the nominal composition and the observed stoichiometry of the synthesized sample is $Co_{3.3}Sn_2S_{1.7}$, which is close to $Co_3Sn_2S_2$.

Fig. 4(a), (b) and (c) show the recorded XPS peaks in Co 2p, Sn 3d and S 2p regions respectively. The observed XPS spectra in all three regions is in agreement with the previously reported results [31]. Fig. 4(a) shows XPS peaks in Co 2p region, showing the peaks of spin orbit doublet of Co viz. $2p_{3/2}$ and $2p_{1/2}$. The peak positions are observed to be at 778.34±0.007eV and 793.27±0.008eV for Co $2p_{3/2}$ and Co $2p_{1/2}$ and are denoted by peaks P1 and P4 in figure 4(a). These values closely match with previously reported ones [31-34]. Co atoms in $Co_3Sn_2S_2$, were considered to be present in metallic Co(0) state similar to Ni atoms in other shandite compound $Ni_3Sn_2S_2$ [35], but the previous reports suggest that there is some charge density around Co atoms due to its bonding with S atoms [32,34]. Here, the observed values of binding energy of XPS peaks in Co 2p region matches with the standard value for metallic Co atoms i.e., 778.3eV [36]. Interestingly, the observed separation between the XPS peaks due to Co $2p_{3/2}$ and Co $2p_{1/2}$ is found to be 14.93eV, which nearly matches with the standard value of 14.97eV [36], in contrast this separation is found to be around 15.3eV for $CoS_2$ [37,38]. The observed value of XPS peaks and separation between spin orbit doublets suggest that Co atoms in $Co_3Sn_2S_2$ forms only metallic bonds with Sn, and exists in Co(0) state. This can also be clarified by the observed value of FWHM of XPS peaks, in our case the FWHM of XPS peaks due to spin orbit doublets of Co 2p orbital is lesser that that for Co dichalcogenides [37-39]. Our results along with the previously reported ones [33] suggest that there Co atoms are present in Co(0) state and the charge of $S^{2-}$ ions is totally compensated by $Sn^{2+}$ ions. Some satellite peaks viz. P2, P3, P4 and P6 are also observed in Co 2p region, which are attributed to the peaks occurred due to surface oxidation of the sample, the occurrence of these satellite peaks is shown in previous reports as well [31,33].

Fig. 4(b) represents the Lorentz fitted XPS peaks in Sn 3d regions showing the presence of peaks due to spin orbit doublet of Sn 3d orbital (Sn $3d_{3/2}$ and Sn $3d_{5/2}$) and peaks of SnO. The observed binding energy for Sn $3d_{5/2}$ and Sn $3d_{3/2}$ are found to be 485.33±0.003eV and 493.73±0.002eV, these values are in agreement with the previously reported values [31-34]. The observed binding energy values are slightly shifted towards higher energies from the standard values [36] for metallic Sn and match with the values obtained for Sn monochalcogenides [40], suggesting Sn to be present in $Sn^{2+}$ state. The separation between the peaks is found to be 8.4eV, which matches with the standard value of 8.41eV [36]. Two other peaks observed in XPS spectra are due to formation of SnO, which occurs due to surface oxidation of the sample.

Fig. 4(c) represents the XPS peaks of synthesized $Co_3Sn_2S_2$ crystal in S 2p region. The observed peaks are due to the spin orbit doublets of S 2p orbital viz. S $2p_{3/2}$ and S $2p_{1/2}$. The binding energy values of S $2p_{3/2}$ and S $2p_{1/2}$ are found to be 162.01±0.008eV and 163.30±0.003eV, these values are consistent with previously reported values [31-34] and shifted from the standard value of S atoms [36]. The shift from the standard value shows S is observed to be towards the lower energy, which suggests covalent bonding between Sn and S and thus the S atoms are present in $S^{2-}$ state in $Co_3Sn_2S_2$. The separation between the two peaks is found to be 1.29eV, which closely matches with the standard value [36]. Our XPS results suggest the chemical states of constituent elements in $Co_3Sn_2S_2$ to be $(Co^0)_3(Sn^{2+})_2(S^{2-})_2$. XPS peaks positions along with their respective FWHM of constituent elements of synthesized $Co_3Sn_2S_2$ single crystal are listed in table 2.

Fig. 5(a) shows magnetization vs temperature (M-T) plot of synthesized $Co_3Sn_2S_2$ single crystal in field cooled (FC) and zero field cooled (ZFC) protocols under the magnetic field of 1kOe. A clear ferromagnetic (FM) transition is observed at around 175K, where the magnetization curve sharply increases in both FC and ZFC measurements. This transition is more pronounced in dM/dT vs T plot, as shown in inset of fig. 5(a), in which a sharp spike can be seen at the transition point. The observed FM transition point ($T_c^{FM}$) is in agreement with the earlier reports. The magnetic moment of synthesized $Co_3Sn_2S_2$ does not saturate right form the FM transition. A baseline has been created in fig. 5(a) to determine the point at which the magnetic moment saturates and a complete FM state is achieved. The magnetic moment is found to saturate below 35K, showing that the spins are completely aligned below 35K and the system attains complete FM state. Interestingly, the magnetization curve in both FC and ZFC measurements coincides to each other, which further shows that the synthesized $Co_3Sn_2S_2$ is homogenously magnetized at 1kOe. This is in contrast with the earlier report [7] in which magnetization measurements were performed on 0.5kOe and a difference between FC and ZFC curves was clearly visible. Fig. 5(b) shows the temperature dependent inverse magnetic susceptibility ($\chi^{-1}$) plot of synthesized $Co_3Sn_2S_2$ single crystal. The magnetic susceptibility in paramagnetic region is found to be well fitted with the Curie Weiss (CW) law, which is as follows:

$$\chi = \frac{C}{T-\theta_p} \qquad (1)$$

Here, C is Curie constant and $\theta_p$ is the Weiss temperature. Curie constant is found to be 4.03emu-mole$^{-1}$-Oe$^{-1}$, while the Weiss temperature ($\theta_p$) is found to be 179K. The observed value of effective magnetic moment ($\mu_{eff}$) can be calculated using the Curie constant as per the following formula

$$\mu_{eff} = \sqrt{\left(\frac{3k_B C}{N_A}\right)} \tag{2}$$

Here, $k_B$ represents Boltzmann constant and $N_A$ is Avogadro number. The calculated value of $\mu_{eff}$ is found to be 5.65$\mu_B$. The value of $\mu_{eff}$ is certainly higher than the theoretical value of Co ions [41,42]. The higher value of $\mu_{eff}$ than the theoretical one, is well studied in literature. The possible reason of the difference between the two values is suggested to be the formation of short-range FM ordering in the paramagnetic state [41,43-45]. Interestingly, Co$_3$Sn$_2$S$_2$ is shown to have short range FM clustering in paramagnetic state in a very report [41] with a strange curvilinear $\chi^{-1}$ vs T plot under the magnetic field of 500Oe above FM transition point. Here, measurements are performed at 1kOe, and curvilinear behaviour is not observed in inverse susceptibility plot. Despite of the difference between inverse susceptibility behaviour, the observed value of $\mu_{eff}$ shows the possible FM clustering above T$_c$, which is more effectively reported in ref. 37. This inhomogeneous magnetic phase is also evident from the different values of FM transition point and Weiss temperature. Weiss temperature is found to be higher than FM transition point, which also indicate the possible presence of short-range magnetic ordering [43].

FM transition involves the change in magnetic entropy at the transition point and the heat capacity measurements are considered to be one of the most reliable methods to probe such transitions. Generally, FM transition appears as sharp spike in the C$_p$ vs T plot. Here, C$_p$ vs T measurements have been carried out and the same is shown in fig. 6. A spike is observed at around 175K showing the presence of phase transition in the synthesized Co$_3$Sn$_2$S$_2$ single crystal. These results are in agreement with the previous report on heat capacity measurements of Co$_3$Sn$_2$S$_2$ [13]. The C$_p$ vs T plot is further fitted with conventional heat capacity equation shown in equation (1), the fitted plot is shown by solid black curve in fig. 6. Generally, heat capacity of a system is contributed by two terms the first one is electronic term (C$_{el}$) and the second one is vibrational term (C$_{vib}$). The total heat capacity is as follows:

$$C_{tot} = C_{el} + C_{vib} \tag{3}$$

Where $C_{el}$ and $C_{vib}$ are given by equation (2) and (3), which are as follows:

$$C_{el} = \gamma T \qquad (4)$$

$$\&\quad C_{vib} = \left(\frac{T}{\theta_D}\right)^3 \int_x^{\frac{\theta_D}{T}} \frac{x^4}{(e^x-1)^2} dx \qquad (5)$$

The observed data is well fitted with the equation (1) and provides the value of Debye temperature, which is found to be 298K. The zoomed view of FM transition in heat capacity measurement along with the fitted plot is shown in inset of fig. 6, which clearly shows that the system undergoes a phase transition (ferromagnetic transition) at 175K.

Fig. 7(a), (b) and (c) show the isothermal magnetization (M-H) plots of synthesized $Co_3Sn_2S_2$ single crystal at 2K, 100K and 250K respectively under the magnetic field range of ±5kOe. The synthesized $Co_3Sn_2S_2$ is found to show ferromagnetic transition at 175K in M-T measurements. M-H plots shows the presence of FM loops temperatures below the transition point viz. 2K and 100K [fig. 7(a) & (b)]. Moment starts to saturate from 0.75kOe at both the temperatures viz. 2K and 100K, which is in agreement with the previous reports [7]. The value of coercivity is found to be 1.4kOe and 1.2kOe at 2K and 100K respectively. The high value of coercivity suggests the presence of hard magnetic phase in synthesized $Co_3Sn_2S_2$ single crystal. The value of saturation magnetization ($M_s$) is found to be 6.3emu/g and 5.6emu/g at 2K and 100K respectively. These values of $M_s$ correspond to 0.54μ$_B$/f.u. and 0.48μ$_B$/f.u. at 2K and 100K respectively. These values are though comparable yet slightly lower than some of the previously reported values [7,9]. M-H plot at 250K is shown in fig. 7(c), and a clear non saturating linear response of magnetization with respect to applied field has been observed. The observed linear M-H plot suggest that the synthesized $Co_3Sn_2S_2$ single crystal is in paramagnetic state at 250K.

Fig. 8(a) shows electrical transport measurements of synthesized $Co_3Sn_2S_2$ single crystal. The resistivity vs temperature (ρ-T) plot shows that the resistivity decreases as the temperature is decreased showing the metallic behaviour of the synthesized $Co_3Sn_2S_2$ single crystal. Here resistivity is measured in both cooling and warming cycle as shown in fig. 8(a). Interestingly, resistivity is not found to follow the same trend during the warming cycle as the same follows during the cooling one. The deviation starts from around 35K and persists upto room temperature. This kind of thermal hysteresis suggest the presence of some sort of charge ordering in the synthesized Co$_3$Sn$_2$S$_2$ single crystal. This type of hysteresis in resistivity is considered to represent the presence of first order phase transition [46]. Interestingly, we do not find any report in literature that show ρ-T measurements in both cooling and warming cycle for Co$_3$Sn$_2$S$_2$ single crystal. Our results are in contrast to ref. 13, in which heat capacity measurements were carried out in both cooling and warming protocols and no hysteresis was present, due to which the observed transition was attributed as a second order phase transition. We repeated the ρ-T measurements on different samples of the same batch maintaining the same cooling and heating rate, and similar hysteresis is observed in each measurement. To get more confidence on observed hysteresis, ρ-T measurements have been carried out in presence of magnetic field of 9T in both cooling and warming cycle maintaining same cooling and heating rate and the same is shown by blue curve in fig. 8(a). A similar hysteresis with significant impact of magnetic field on resistivity value is observed in ρ-T measurements at 9T. This hysteresis is observed in the region, in which the spins are not completely aligned and a full FM state is not achieved. The observed behaviour is well supported by M-T measurement, which shows that the complete FM state occurs at below 35K. The observed hysteresis in ρ-T measurements creates the possibility that the observed FM transition may be of first order. The ρ-T data is fitted in complete temperature range (300K-2K) using the Bloch-Grüneisen (B-G) formula and the fitted plot is shown as the solid black curve in inset of figure 8(b). According to B-G formula, ρ(T) is well described by the following formalism,

$$\rho(T) = \left[\frac{1}{\rho_s} + \frac{1}{\rho_i(T)}\right]^{-1} \tag{6}$$

Here, $\rho_s$ denotes saturation resistivity which is independent of temperature and $\rho_i(T)$ is given by following equation,

$$\rho_i(T) = \rho(0) + \rho_{el\text{-}ph}(T) \tag{7}$$

here ρ(0) represents residual resistivity arising due to impurity scattering, the second term $\rho_{e\text{-}ph}(T)$ represents temperature dependent term, which depend on electron-phonon scattering. Further, $\rho_{e\text{-}ph}(T)$ is given by the following formula

$$\rho_{el-ph} = \alpha_{el-ph}\left(\frac{T}{\theta_D}\right)^n \int_0^{\frac{\theta_D}{T}} \frac{x^n}{(1-e^{-x})*(e^x-1)} dx \tag{8}$$

here $\alpha_{el-ph}$ is electron-phonon coupling parameter, $\theta_D$ represents Debye temperature and n is constant. The value of n determines the dominant scattering in the studied system, here the ρ-T data is well fitted with the above equation for n=5, signifying dominant electron-phonon scattering. Residual resistivity ratio (RRR) is found to be 21, which shows high metallicity of the synthesized crystal. The observed RRR value is comparable to the previously reported values [7,9,21]. The obtained value of Debye temperature $\theta_D$ from above fitting formula is 302K, which is comparable to the value obtained from fitted heat capacity plot.

Fig. 8(b) is showing ρ-T-H plots of synthesized $Co_3Sn_2S_2$ single crystal under zero magnetic field and magnetic field of 1T, 5T, 9T and 14T. All the measurements are performed in cooling cycle. The kink that is observed in zero field resistivity plot is disappeared in ρ-T plots at the magnetic field above 5T. A comparison has been made between the ρ-T plots at zero magnetic field and under the magnetic field of 14T, and an interesting phenomenon is observed at temperatures below the FM transition point i.e., 175K. Zero field resistivity is higher from the same at 14T below the transition point upto 70K, showing the presence of negative magnetoresistance (MR). The resistivity value is higher at 14T than zero field resistivity below 70K, showing the presence of positive MR. MR% vs T plot has been drawn to more elucidate the observed results and the same is shown in inset of fig. 7(b). MR% has been calculated by the using the following formula:

$$MR\% = \left[\frac{\rho(H)-\rho(0)}{\rho(0)}\right] \times 100 \qquad (9)$$

Here, ρ(H) is field dependent resistivity and ρ(0) is zero field resistivity. A zero-base line has been created to check at which temperature the MR% turns out to be negative from the positive value. A high positive MR% is seen at temperature 50K, which is due to topological non trivial character of $Co_3Sn_2S_2$, or this MR% can be identified as surface states dominated one. At temperature above 70K, the observed MR% value turns out to be negative and the same persists upto 175K, which is FM transition point of synthesized $Co_3Sn_2S_2$ single crystal. This negative MR% is dominated by the ferromagnetic properties of the synthesized $Co_3Sn_2S_2$ single crystal. At temperature above 175K, the sample is in paramagnetic state and again a small but positive MR% has been observed. For further clarification of the observed behaviour, MR% has been calculated at different temperatures in all three regions and the same is shown in fig. 8(c) and 8(d).

Fig. 8(c) shows the MR% vs H plot in positive MR region i.e., at temperatures 2K, 5K, 10K, 20K, 30K, 50K, 70K. The synthesized $Co_3Sn_2S_2$ is found to show unsaturated high MR around 220% at 2K, 5K, 10K. MR% remain intact at these three temperatures. The observed MR% at 2K is in agreement with the previous report [7,21]. MR% is found to decrease as the temperature is increased and a small positive MR% of 6% is observed at 70K. The suppression of MR% with temperature shows that the bulk conducting channels starts to dominate in conduction mechanism as the temperature is increased. The synthesized $Co_3Sn_2S_2$ single crystal show negative MR at temperature above 70K as can be seen in fig. 8(d), which depicts the MR% vs H plot at 100K, 125K, 150K and 250K. The observed negative MR at these temperatures is observed in some earlier reports too [7,30]. This negative MR is found to be linear and increases as the temperature is increased. Generally, the ferromagnetic materials show a linear negative magnetoresistance caused by suppressed spin scattering due to its magnetic state [47,48]. The observed negative MR is clearly related to the ferromagnetic state of $Co_3Sn_2S_2$, as the same turns out to be positive when the temperature is raised above transition temperature. The synthesized $Co_3Sn_2S_2$ is found to show a small quadratic MR in its paramagnetic state due to dominant Lorentz deflection mechanism. Further, the conduction mechanism in synthesized $Co_3Sn_2S_2$ single crystal is analysed by fitting the positive MR% at low temperatures with power law i.e., MR%=A*$H^y$, and the same is shown in fig 8(e). The shape of MR% provides important information about the origin of MR in topological materials. $Co_3Sn_2S_2$ is reported to have quadratic or parabolic MR in previous reports, which arises due to change in trajectory of charge carriers in presence of magnetic field [7,30]. The quadratic MR is considered as classical MR as the same has classical origin. Here, the observed MR at low temperatures is neither perfectly parabolic nor it seems to be linear as observed for Dirac semimetals. Here, to elucidate the deviation of MR% from linear MR, MR% data at 2K, 20K, 30K, 50K and 70K are fitted with power law and shown in fig. 8(e). The value of exponent $\gamma$ determines whether the conduction is dominated by bulk states or the topological surface states. The value of $\gamma$ near to 1 (linear) shows solely surface states dominated conduction mechanism while the value of $\gamma$ near to 2 (quadratic) shows the complete suppression of topological surface states. An intermediate value of $\gamma$ shows that the both viz. topological surface states and bulk states are simultaneously contributing in conduction mechanism. For synthesized $Co_3Sn_2S_2$ single crystal, the value of $\gamma$ is found to be 1.31 at 2K, which is an intermediate value of 1 and 2 and shows that the conduction mechanism is contributed by both topological surface state and bulk conducting state [49]. The value of exponent γ tends to increase as the temperature is increased as shown in fig. 8(f). The increment in the value of γ with temperature shows that the bulk contribution in conduction mechanism is increasing with temperature. The value of γ is found to be 2.06 at 70K, showing perfectly parabolic nature of MR at 70K. These results suggest that the contribution of surface states in conduction mechanism is completely suppressed at 70K, and the bulk magnetic states of synthesized $Co_3Sn_2S_2$ single crystal dominate in conduction mechanism at temperatures above 70K. These results are in agreement with field dependent ρ-T measurements.

Some evidence of contribution of topological surface states in transport phenomenon are observed in MR measurements, this invokes us to analyse the low field magnetoconductivity in context of possible weak antilocalization (WAL) effect in synthesized $Co_3Sn_2S_2$ single crystal. Fig. 9(a),(b),(c),(d)&(e) are showing the fitted magnetoconductivity (MC) of synthesized $Co_3Sn_2S_2$ single crystal in a field range of ±1T at 2K, 5K, 10K, 20K and 30K. Presence of WAL effect in a topological material, can be confirmed by fitting low temperature MC with HLN formalism [50]. In HLN formalism, the difference in magnetoconductivity Δσ(H) is modelled with the following equation:

$$\Delta\sigma(H) = -\frac{\alpha e^2}{\pi h}\left[\ln\left(\frac{B_\varphi}{H}\right) - \Psi\left(\frac{1}{2} + \frac{B_\varphi}{H}\right)\right] \quad (10)$$

Here, $B_\phi$ is the characteristic field and is given by $B_\varphi = \frac{h}{8e\pi l_\varphi^2}$, $l_\phi$ is phase coherence length and Ψ is digamma function. The phase coherence length $l_\phi$ is defined as the maximum length travelled by electron while maintaining its phase. The value of pre-factor α determines the nature of localization present in the system. A positive value of α signifies weak localization (WL) effect, while a negative value of α indicates the presence of WAL effect. In topological materials, WAL is supposed to be induced by topological surface states. These topological surface states are protected by π Berry phase. A material with π Berry phase, takes the value of α to be -0.5 per conducting channel [51]. If α takes the value to be equal to -1, it shows that there are two distinct TSS are present in the system, one is at the top and other at the bottom [50,52]. The deviation in value of α from standard values viz. -0.5 and -1, shows that the top and bottom surface states are connected through bulk conducting channels [53]. Here, the obtained value of pre-factor α and phase coherence length $l_\phi$ at different temperatures are listed in table-3. The value of α at 2K is found to be -0.48, which is near to -0.5, and suggest that single TSS contributes to the conduction mechanism in synthesized $Co_3Sn_2S_2$ single crystal at 2K. Also, the value of α appears to be independent of temperature upto 10K, and the same is found to decrease as the temperature is increased as shown by blue encircled dots in Fig. 9(f). The value of α is found to be -0.37 at 30K. In topological materials, the observed WAL effect is explained by considering one or two surface bands that results with the values of α being either -0.5 or -1 [54]. The deviation of the value of α from the standard values viz. -0.5, suggest that the conduction mechanism is contributed by multiple conducting channels, which includes topological surface states and bulk conducting states. It has been seen in various reports that the value of α decreases from the standard values i.e., -0.5 (single TSS) and -1 (two distinct TSS) [54-57]. Here, the deviation in value of α form -0.5 is observed at temperatures above 10K, which signifies multiple channels contribution in conduction mechanism above 10K.

In fig. 9(f), the variation of inverse of square of phase coherence length $l_\phi^{-2}$ with respect to temperature is shown by red symbols. Temperature dependence of $l_\phi$ in a system is important to study the scattering processes and dephasing mechanism. According to Nyquist theory, linear relationship of $l_\phi^{-2}$ with temperature shows the presence of only single electron-electron (e-e) scattering [58,59]. Here, it is clear from fig. 9(f) that $l_\phi^{-2}$ vs T plot is not linear, suggesting the presence of two scattering processes viz. e-e and e-p. At low temperatures the dominant scattering is the e-e scattering while the e-p scattering occurs at higher temperatures due to contribution of bulk conducting states [59-61]. The 2D and 3D nature of WAL effect can also be determined by fitting the $l_\phi^{-2}$-T plot with the following power law [60]:

$$\frac{1}{l_\Phi^2(T)} = \frac{1}{l_\Phi^2(0)} + A_{e-e}T^p + A_{e-p}T^q \tag{11}$$

Here, $l_\phi(0)$ is the dephasing length at absolute zero and $A_{e-e}T^p$ and $A_{e-p}T^q$ represents the contribution from e-e scattering and e-p scattering respectively. The fitted plot is shown with solid black line in fig. 9(f). Here, the value of p and q are found to be 1 and 2 respectively and the value of $l_\phi(0)$ is found to be 50nm. The values of $A_{e-e}$ and $A_{e-p}$ are found to be $2.12\times10^{-6}$ $nm^{-2}$-$K^{-1}$ and $1.50\times10^{-7}$ $nm^{-2}$-$K^{-2}$ respectively. These results show that both the scattering process viz. e-e and e-p scattering play role in dephasing mechanism of $Co_3Sn_2S_2$. The systems having only 2D e-e scattering shows the linear relationship between $l_\phi^{-2}$ and T [59]. If the e-p scattering is present in the system, the power law is represented as $l_\phi^{-2} \propto T^x$ [59,61,62], where x determines dimensionality of present e-p coupling. For 3D e-p coupling, x takes the value 3 and the same takes the value equal to 2 for 2D e-p coupling [61,62]. Here, power law fitted results shows that the exponent of e-p coupling term i.e., q is equal to 2, showing the presence of 2D e-p coupling. Also, the presence of 2D e-e and e-p scattering in synthesized $Co_3Sn_2S_2$ single crystal suggest that the 2D conducting channels contribute in the observed WAL effect.

In literature, we did not find any report on WAL effect in $Co_3Sn_2S_2$ single crystals, which raises question whether the observed WAL effect is intrinsic behavior of the synthesized $Co_3Sn_2S_2$ single crystal or any impurity in the sample results in occurrence of WAL effect. To clarify this issue, we performed angle dependent magneto transport measurements maintaining constant temperature at 2K. Also, the angle dependent magneto transport measurements are crucial to confirm that whether the observed WAL effect arises due to 2D topological surface states or bulk conducting states are responsible for occurrence of the same [51]. The geometry used for angle dependent magneto transport measurements is shown in inset of fig. 10(a). The current is flowing along x-axis while the magnetic field is applied along z axis. The angle ($\theta$) is the tilt angle measured from the z axis. Fig. 10(a) exhibit MR% vs applied magnetic field plot at various angles viz. 0°, 15°, 45° and 90°, and the temperature remained constant at 2K throughout the measurement. MR% showing to have anisotropic behavior as MR% gradually decreases as the angle is increased. This shows that the observed results in magneto transport measurements depict intrinsic properties of synthesized $Co_3Sn_2S_2$ single crystal, as any impurity induced phenomenon is not supposed to show such kind of anisotropy. The observed behavior in angle dependent transport measurements hints towards topological surface states dominated transport phenomenon as the same is strongly depends on the direction of field. Also, the normalized conductivity has been plotted against Hcos$\theta$ and the same is shown in Fig. 10(b). Here, all three plots are found to be merged in a single plot at low magnetic field values and are deviated at higher magnetic fields. This shows that the observed WAL effect is mainly contributed by 2D topological surface states of synthesized $Co_3Sn_2S_2$ single crystal. This is considered to be the most common and effective approach to determine the presence of 2D WAL effect in topological materials [51,53,63-66].

**Conclusion:**

Summarily, single crystals of $Co_3Sn_2S_2$ are synthesized using a 3-step method based on solid state reaction route. The single crystal is thoroughly characterized in context of phase purity and crystallinity. XPS measurements confirms the presence of metallic bonding between Co and Sn atoms forming kagome layers. Ferromagnetic transition at 175K is clearly observed in magnetization, heat capacity and electrical transport measurement. Large positive MR is observed at low temperatures, which gradually decreases and turns out to be negative above 70K. The MR is found to be strongly dependent on bulk magnetic state of $Co_3Sn_2S_2$ single crystal above 70K. Clear signatures of the contribution of topological surface states are observed in MR, which further lead to presence of WAL effect. The HLN fitted MC of synthesized $Co_3Sn_2S_2$ shows that WAL effect is present in the system at low temperatures.

**Acknowledgement:**

Authors are thankful to Director of National Physical Laboratory for his encouragement and keen interest in research activities. Authors are thankful to Dr. Pallavi Kushwaha for magnetization measurements on QD-MPMS. Authors would like to thank Dr. J. S. Tawale for FESEM measurements. Kapil Kumar is thankful to UGC, India for research fellowship. M. M. Sharma is thankful to CSIR, India for research fellowship. Kapil Kumar and M. M. Sharma are thankful to AcSIR, India for Ph.D. registration.

## Table-1

Parameters obtained from Rietveld refinement:

| Cell Parameters | Refinement Parameters |
|---|---|
| Cell type: Rhombohedral<br>Space Group: R -3 m<br>Lattice parameters: a=b=5.376(1)Å<br>& c=13.155(3)Å<br>$\alpha=\beta=90°$ & $\gamma=120°$<br>Cell volume: 380.197Å$^3$<br>Density: 21.774g/cm$^3$<br>Atomic co-ordinates:<br>Co (0,0.5,0.5)<br>Sn1 (0,0,0.5)<br>Sn2 (0,0,0)<br>S (0,0,0.2805) | $\chi^2$=1.79<br>R$_p$=6.14<br>R$_{wp}$=7.93<br>R$_{exp}$=5.87 |

## Table-2

XPS peaks position and FWHM of constituent elements of synthesized $Co_3Sn_2S_2$ single crystal:

| Element | Spin-orbit doublet | Binding Energy | FWHM |
|---|---|---|---|
| Sn | 3d$_{5/2}$ | 485.33±0.003eV | 0.84±0.02eV |
|  | 3d$_{3/2}$ | 493.73±0.002eV | 0.69±0.03eV |
| Co | 2p$_{3/2}$ | 778.34±0.007eV | 0.90±0.02eV |
|  | 2p$_{1/2}$ | 793.27±0.008eV | 1.69±0.15eV |
| S | 2p$_{3/2}$ | 162.01±0.008eV | 1.15±0.05eV |
|  | 2p$_{1/2}$ | 163.30±0.003eV | 1.16±0.010eV |

## Table: 3

Low field (up to ±1 Tesla) HLN fitted parameters of $Co_3Sn_2S_2$ single crystal

| Temperature(K) | $\alpha$ | l$_\phi$ |
|---|---|---|
| 2 | -0.48(8) | 48.93nm |
| 5 | -0.49(8) | 49.60nm |
| 10 | -0.48(2) | 48.74nm |
| 20 | -0.39(6) | 43.83nm |
| 30 | -0.37(7) | 41.03nm |

**Figure captions:**

**Figure 1:** Schematic of heat treatment followed to synthesize $Co_3Sn_2S_2$ single crystal.

**Figure 2: (a)** XRD pattern taken on crystal flake of synthesized $Co_3Sn_2S_2$ single crystal. **(b)** Rietveld refined XRD pattern of $Co_3Sn_2S_2$ taken on crushed powder. **(c)** Unit cell structure of synthesized $Co_3Sn_2S_2$ single crystal processed in VESTA software.

**Figure 3: (a)** FESEM image of surface morphology of synthesized $Co_3Sn_2S_2$ single crystal. **(b)** EDS spectra of synthesized $Co_3Sn_2S_2$ single crystal, in which inset is showing the elemental composition of constituent elements of $Co_3Sn_2S_2$.

**Figure 4:** XPS spectra of synthesized $Co_3Sn_2S_2$ single crystal recorded in **(a)** Co- 2p region **(b)** Sn-3d region **(c)** S-2p region.

**Figure 5: (a)** M-T plot of synthesized $Co_3Sn_2S_2$ single crystal under a magnetic field of 1 kOe, in which inset is showing the dM/dT vs T plot of the same showing the presence of FM transition. **(b)** CW fitted inverse susceptibility ($\chi^{-1}$) vs T plot under the magnetic field of 1kOe.

**Figure 6:** Fitted $C_p$ vs T plot is synthesized $Co_3Sn_2S_2$ single crystal, in which inset is showing the zoomed view of the same around FM transition.

**Figure 7:** Isothermal M-H plots of synthesized $Co_3Sn_2S_2$ single crystal at different temperatures viz. **(a)** 2 K **(b)** 100 K and **(c)** 250 K.

**Figure 8: (a)** $\rho$-T-H plots of synthesized $Co_3Sn_2S_2$ single crystal in cooling and warming cycle under the magnetic field of 0T and 9T, in which inset is showing the B-G fitted $\rho$-T plot of the same taken in cooling cycle **(b)** $\rho$-T plot in zero field and under the magnetic field of 14T, 9T, 5T, 1T and zero field, in which inset is showing the MR% vs temperature plot of the same at 14T. **(c)** MR% vs H plot of synthesized $Co_3Sn_2S_2$ single crystal at 2 K, 5 K, 10 K, 20 K, 30 K, 50 K and 70 K. **(d)** MR% vs H plot of synthesized $Co_3Sn_2S_2$ single crystal at 100 K, 125 K, 150 K and 250 K. **(e)** Power law fitted MR% vs H plot of synthesized $Co_3Sn_2S_2$ single crystal in positive magnetic field at 2 K, 10 K, 20 K, 30 K, 50 K and 70 K. **(f)** The variation of exponent '$\gamma$' with respect to temperature.

**Figure 9:** HLN fitted low field ($\pm$ 1 T) MC of synthesized $Co_3Sn_2S_2$ single crystal at **(a)** 2 K **(b)** 5 K **(c)** 10 K **(d)** 20 K **(e)** 30 K **(f)** The variation of HLN fitting parameters with respect to temperature.

**Figure 10: (a)** MR% vs H plots at various tilt angles viz. 0°, 15°, 45°, 90° and temperature is maintained at 2K, in which the inset is showing the geometry used to perform angle dependent transport measurements. **(b)** Normalized $\sigma$ vs Hcos$\theta$ plot of synthesized $Co_3Sn_2S_2$ single crystal measured at 2K with different tilt angles viz. 0°, 15°, 45°.

Fig. 1:

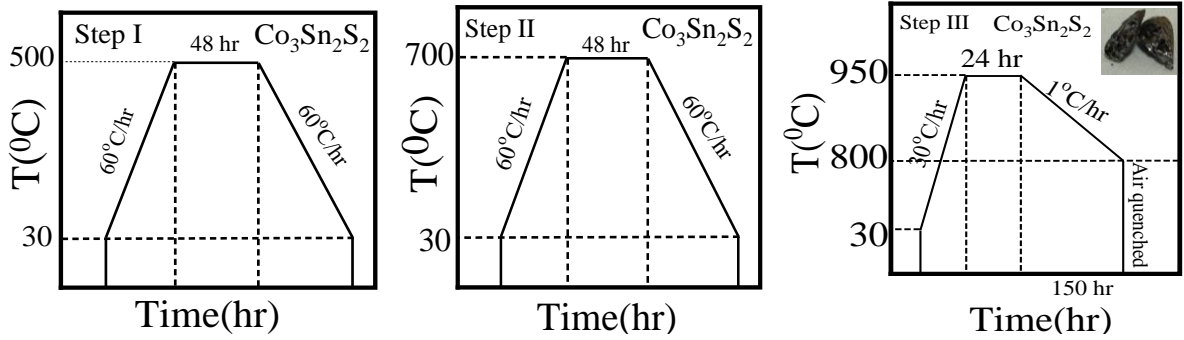

Fig. 2(a):

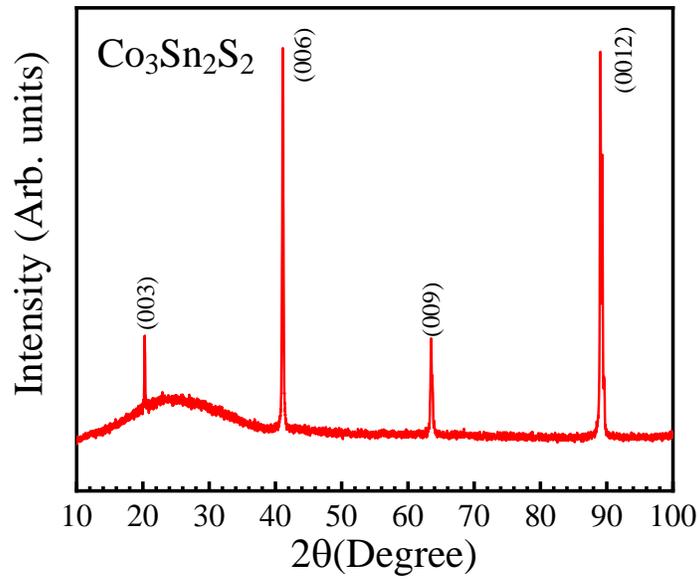

Fig. 2(b)&(c)

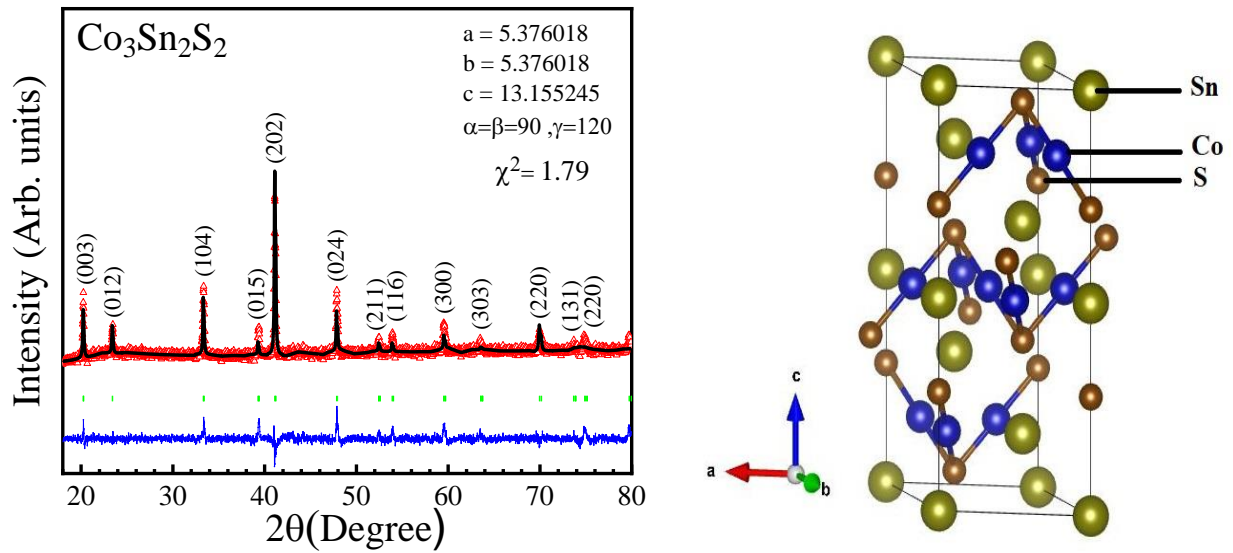

Fig.3

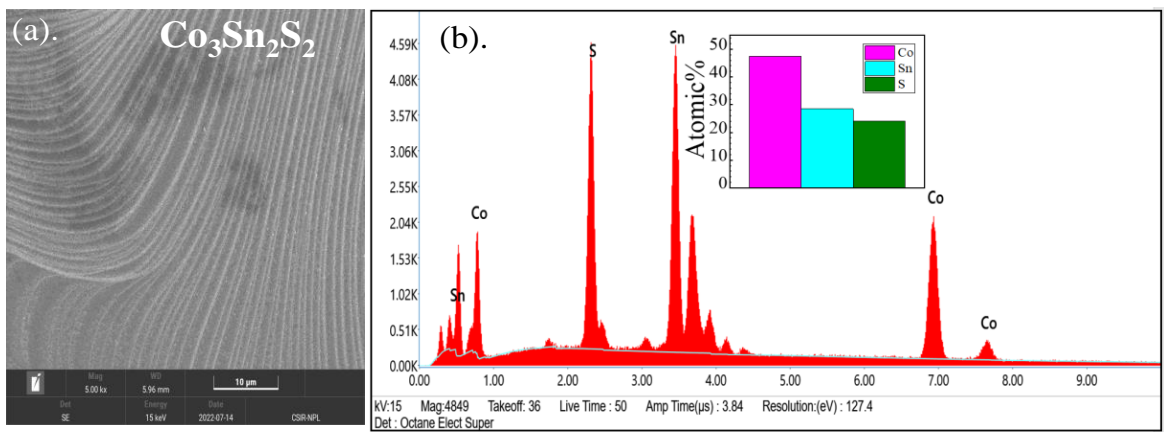

Fig. 4

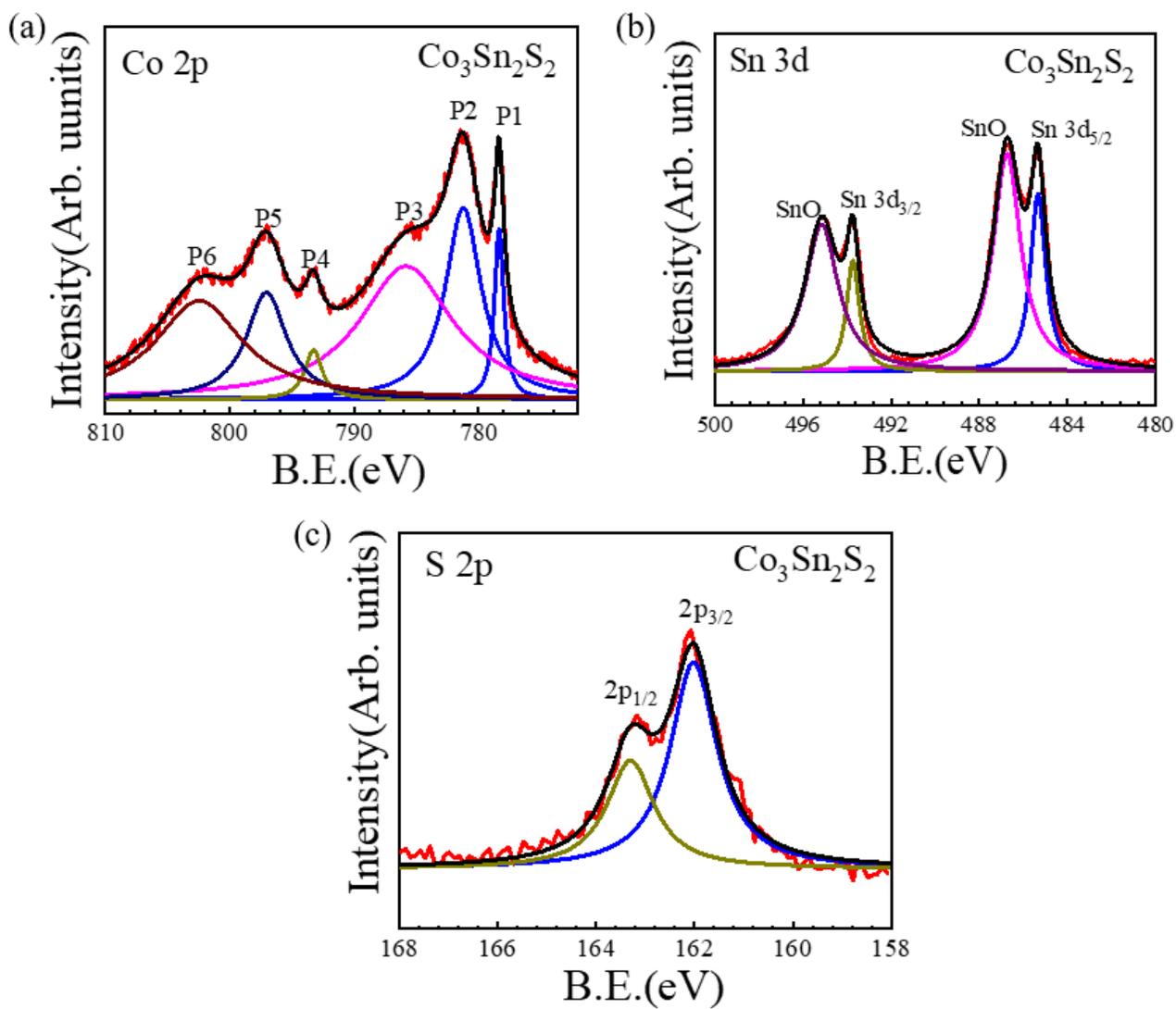

Fig.5

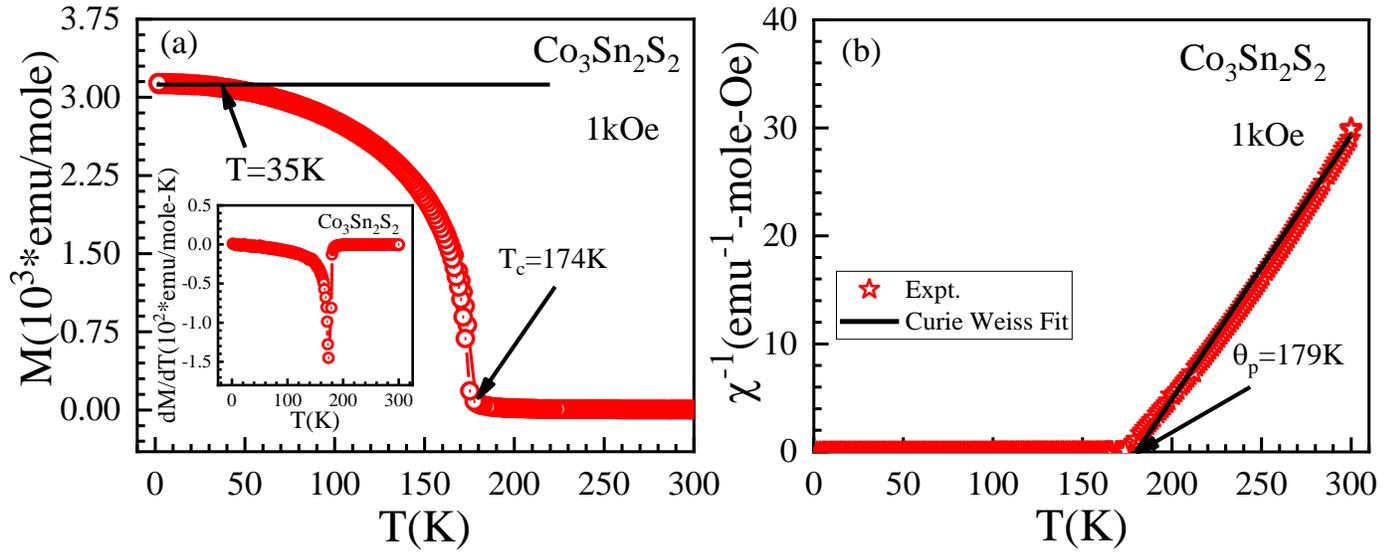

Fig. 6

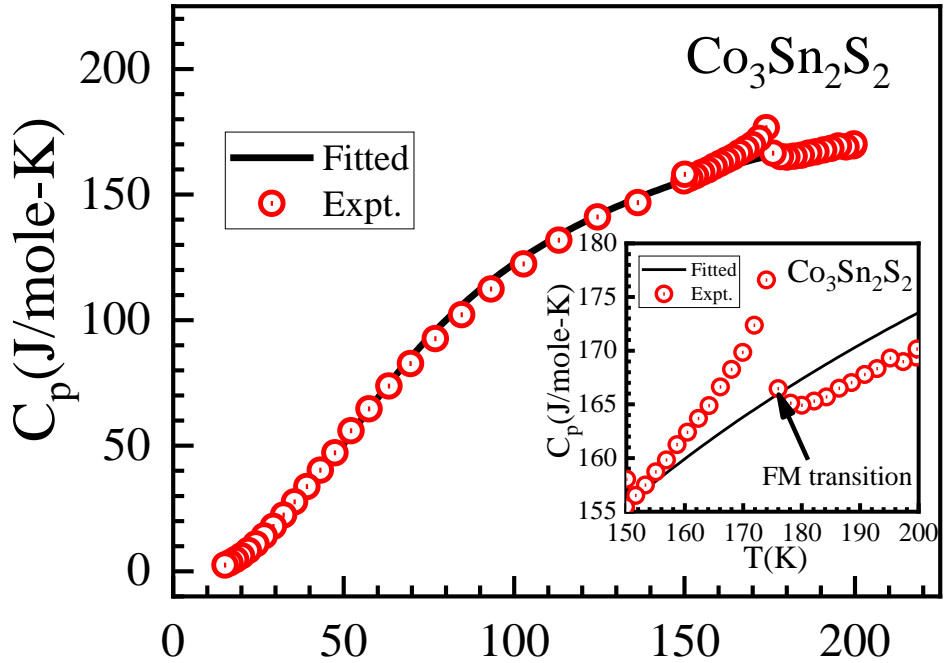

Fig. 7

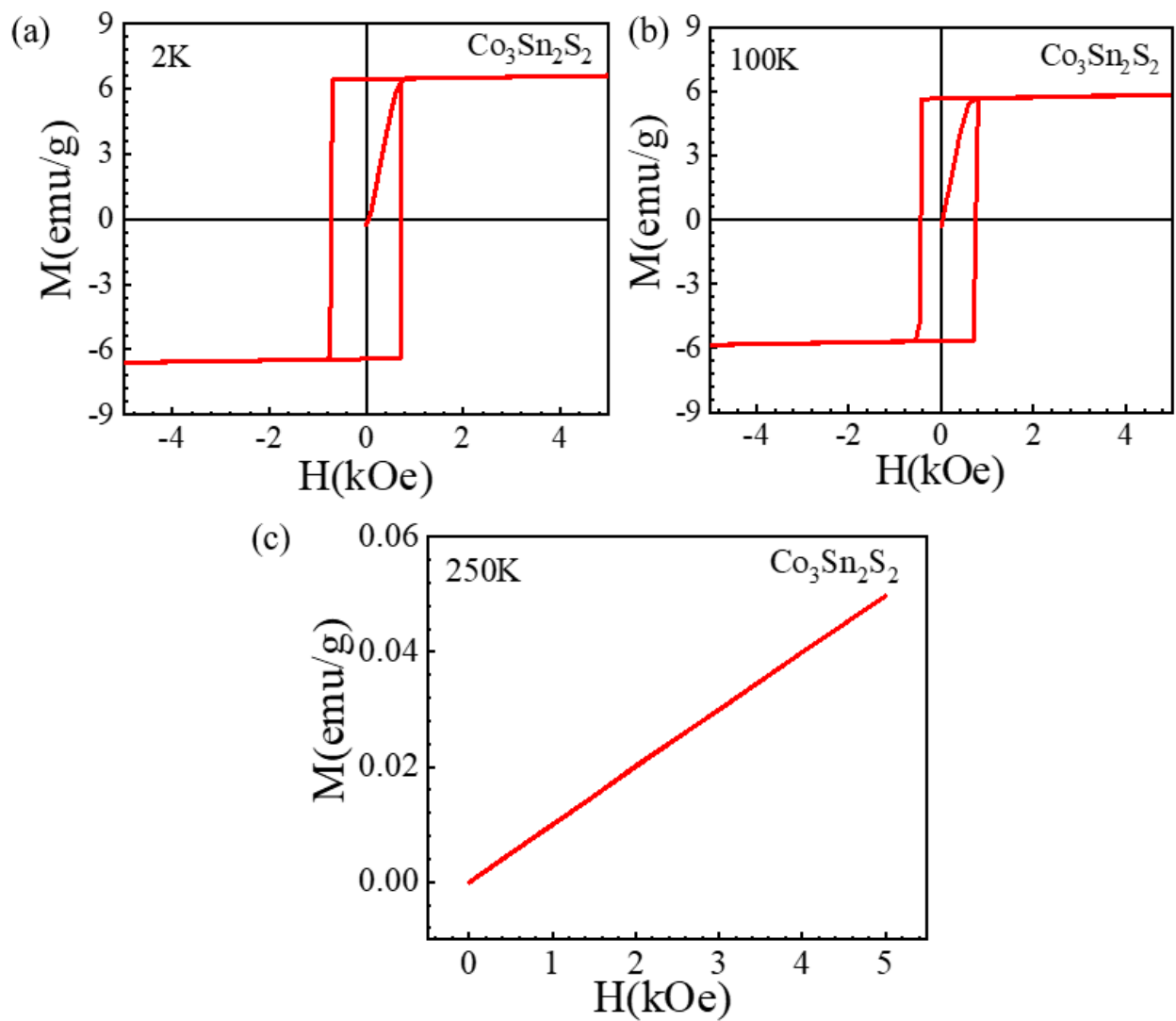

Fig. 8

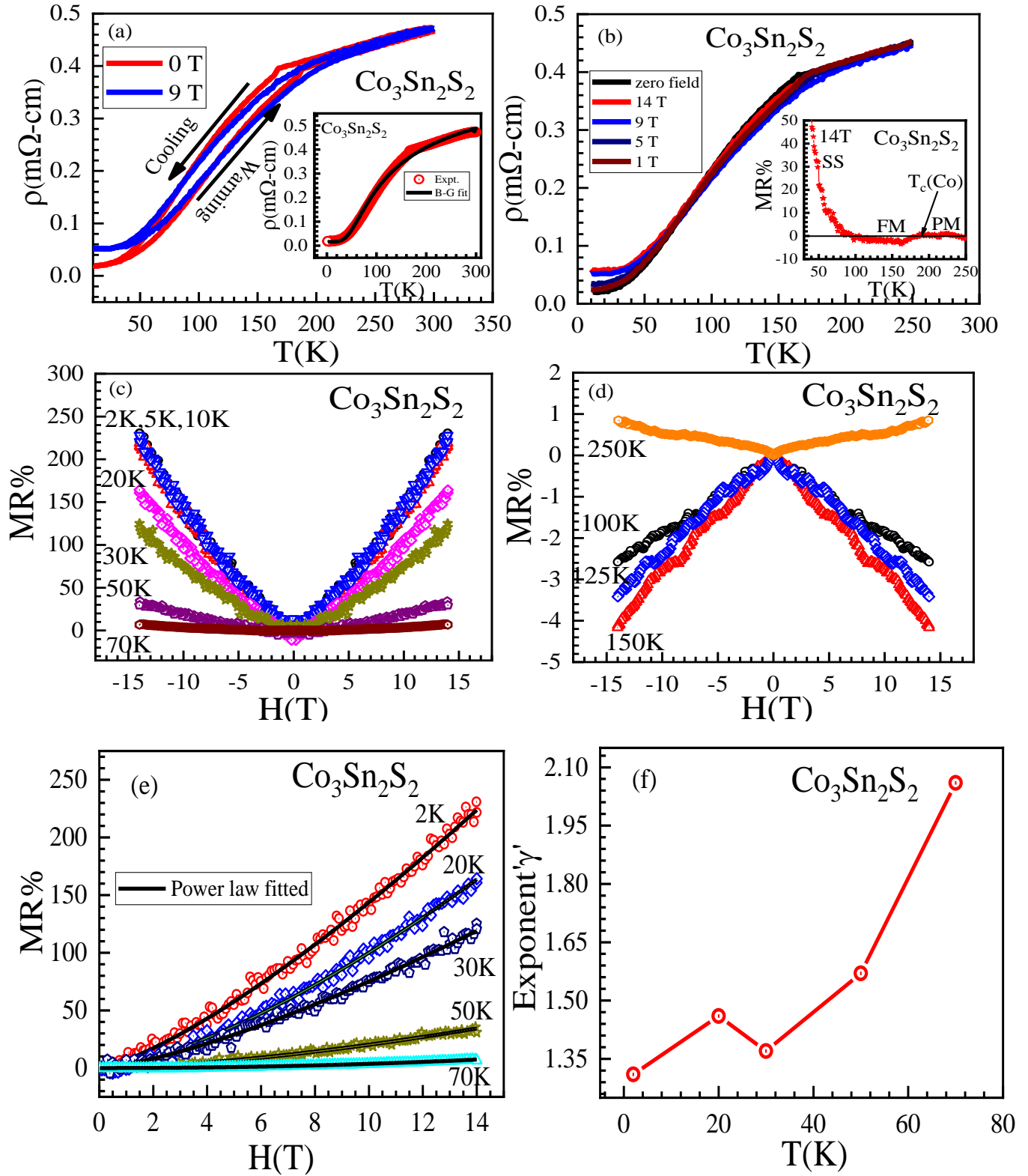

Fig. 9

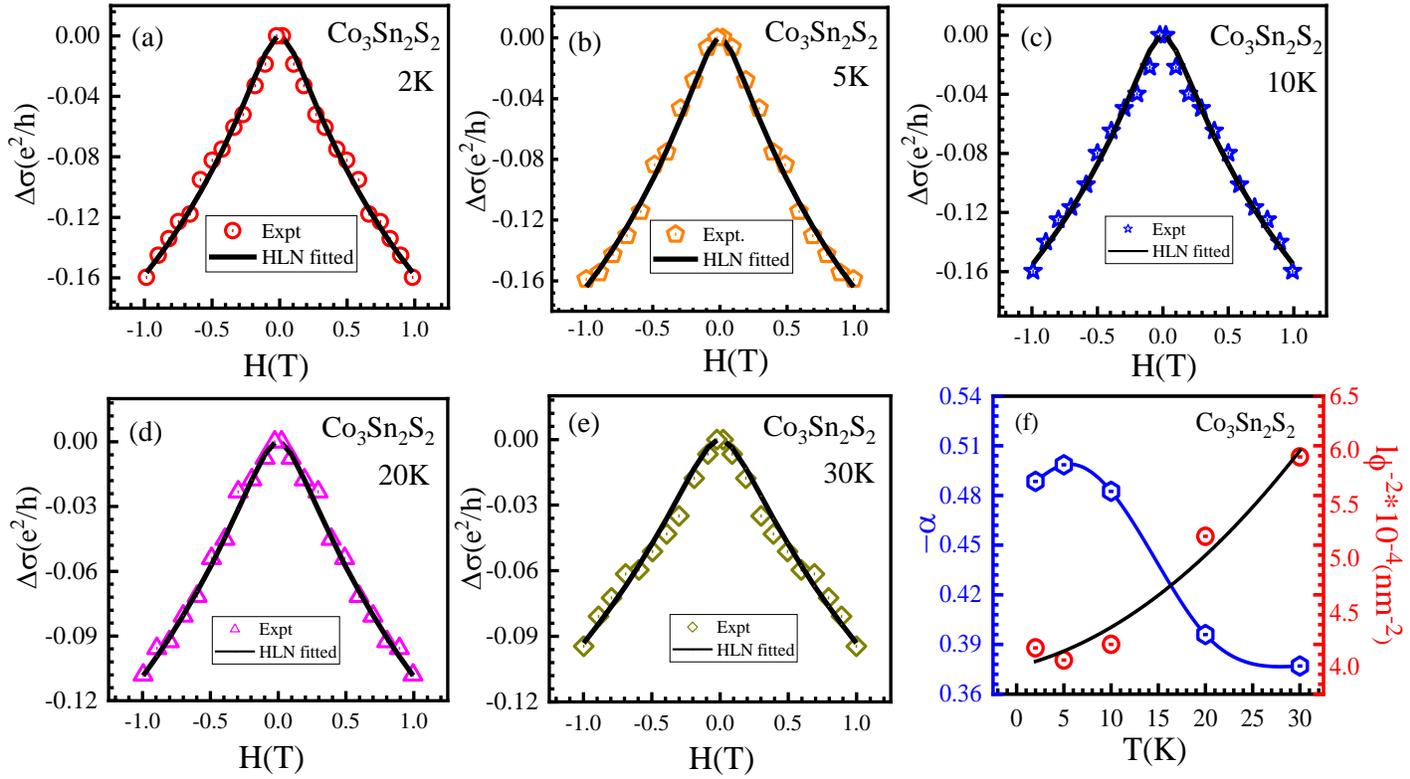

Fig. 10

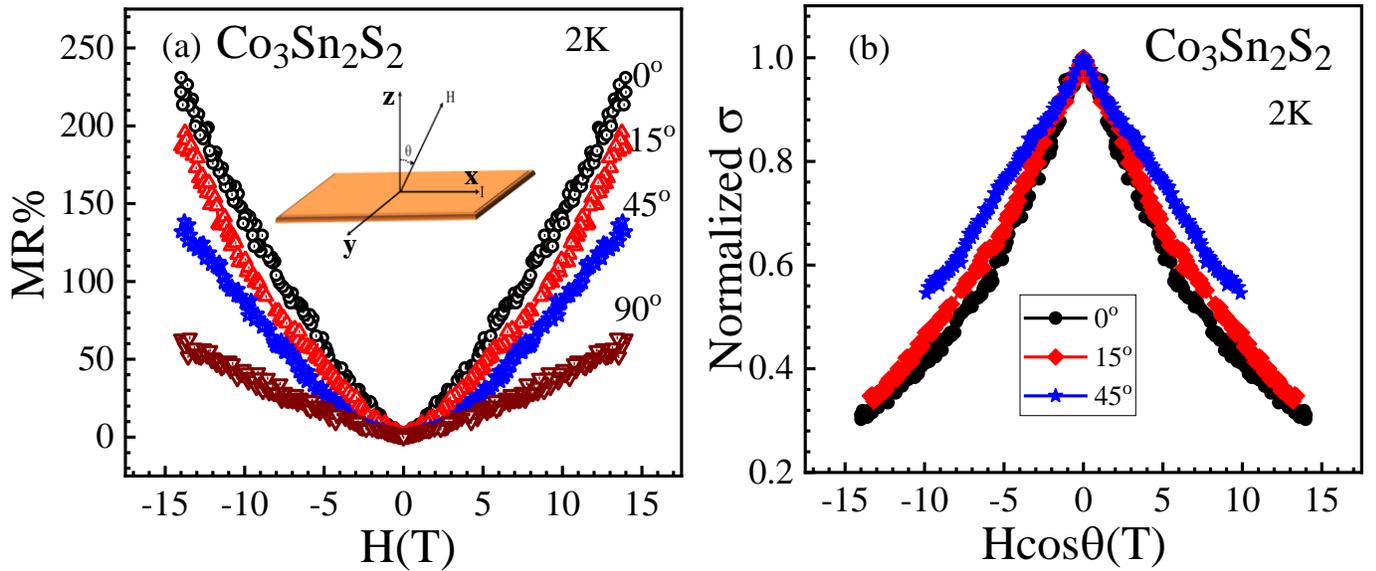